\documentstyle[psfig, amssym, 11pt, aaspp4, flushrt]{article}

\slugcomment{Accepted for publication in Astrophysical Journal}

\lefthead{Chadwick et al.}

\righthead{VHE Gamma Rays from PKS 2155--304}

\begin{document}

\title{VHE Gamma Rays from PKS 2155--304} 

\author{P.~M.~Chadwick, K.~Lyons, T.~J.~L.~McComb, K.~J.~Orford,
J.~L.~Osborne, S.~M.~Rayner, S.~E.~Shaw, K.~E.~Turver, and
G.~J.~Wieczorek}

\affil{Department of Physics, Rochester Building, Science Laboratories,
University of Durham, Durham, DH1 3LE, U.K.}

\authoremail{k.e.turver@dur.ac.uk}

\begin{abstract}

The close X-ray selected BL Lac PKS 2155--304 has been observed using
the University of Durham Mark 6 very high energy (VHE) gamma ray
telescope during 1996 September/October/November and 1997
October/November. VHE gamma rays with energy $ > 300$ GeV were detected
from this object with a time-averaged integral flux of $(4.2 \pm
0.7_{stat} \pm 2.0_{sys}) \times 10^{-11}~{\rm cm}^{-2}~{\rm s}^{-1}$.
There is evidence for VHE gamma ray emission during our observations in
1996 September and 1997 October/November, with the strongest emission
being detected in 1997 November, when the object was producing the
largest flux ever recorded in high-energy X-rays and was detected in $>
100$ MeV gamma-rays. The VHE and X-ray fluxes show evidence of a
correlation.

\end{abstract}

\keywords{BL Lacertae objects: individual (PKS 2155--304) --- galaxies:
active --- gamma rays: observations}

\section{Introduction}

Close X-ray selected BL Lacs (XBLs) are sources of very high energy
(VHE) gamma rays at energies above 300 GeV. The BL Lac first detected as
a source of VHE gamma rays was Mrk 421 (\cite{punch1992}), following its
discovery as a GeV source by the EGRET experiment on the {\it Compton
Gamma Ray Observatory} (\cite{lin1992}). Mrk 421 has been extensively
monitored in VHE gamma rays and exhibits complex behaviour, having a
low-level quiescent state with flaring on time scales as short as 30
minutes (\cite{gaidos1996}; \cite{zweerink1997}; \cite{takahashi1998}).
Mrk 501 is also a source of VHE gamma rays (\cite{quinn1996}), although
not detected at GeV energies with EGRET, and also exhibits low-level
emission with flaring (\cite{catanese1997a}). In 1997, extremely strong
outbursts of TeV emission were detected lasting several months
(\cite{deckers1997}, \cite{quinn1997}, \cite{hayashida1998}) The BL Lac
1ES 2344+514 also emits VHE gamma rays, most of the evidence for
emission coming from a single outburst (\cite{catanese1997b},
\cite{catanese1998}). All of these objects are close ($z \sim 0.03$),
X-ray selected BL Lacs. There have been no reported detections of VHE
gamma rays from radio-selected BL Lacs (RBLs --- see e.g.
\cite{kerrick1995}, \cite{roberts1998}) although this category of object
is frequently detected at GeV energies by the EGRET experiment.

\cite{stecker1996} have interpreted the gamma ray results in the GeV --
TeV range and propose a model in which RBLs will be GeV gamma ray
sources and XBLs will be TeV sources. They associate the emission of TeV
$\gamma$-rays from XBLs with the presence of relativistic electrons with
energies higher than those in RBLs. They then go on to show that a
simple synchrotron self-Compton model can explain the differences
observed between RBLs and XBLs provided the attenuation of the VHE gamma
ray flux by pair production with the intergalactic infrared background
is taken into account (\cite{stecker1992}; \cite{stecker1997}). On the
basis of this model, PKS 2155--304, despite having a redshift of 0.117,
is predicted to be a strong TeV gamma-ray source.

The BL Lac PKS 2155--304 was discovered as an X-ray source during
observations made with the {\it HEAO-1} satellite (\cite{schwartz1979};
\cite{griffiths1979}) at a position where the {\it Ariel~V} satellite
had previously detected confused emission (\cite{cooke1978}). In many
ways, PKS 2155--304 may be regarded as the archetypal X-ray selected BL
Lac object; like most BL Lac objects it is associated with a compact,
flat spectrum radio source and has an almost featureless continuum which
extends from radio to X-ray energies. It is the brightest known BL Lac
at UV wavelengths, and the object's maximum power is emitted between the
UV and the soft X-ray range (\cite{wandel1991}). PKS 2155--304 has a
history of rapid, strong broadband variability and has been the subject
of several multiwavelength monitoring campaigns (see e.g.
\cite{brinkmann1994}; \cite{courvoisier1995}; \cite{pesce1997}). In 1997
November, contemporaneous with some of the observations reported here,
X-ray emission was detected with the {\it Beppo-SAX} satellite
(\cite{chiappetti1997}) with a flux equal to the strongest previous
outburst.

The EGRET experiment on board the {\it Compton Gamma Ray Observatory}
was used to detect 30 MeV -- 10 GeV gamma ray emission from PKS
2155--304 during 1994 November 15 -- 29 (\cite{vestrand1996}). These
observations indicated a very hard spectrum, with an integral power-law
spectral index of $1.71 \pm 0.24$, and this, combined with its
proximity, make it an excellent candidate TeV source. However, a recent
detailed analysis of the shape of the X-ray spectrum suggests that PKS
2155--304 may not be an important TeV emitter, producing photons with a
maximum energy of $\sim 800$ GeV, although a very high degree of beaming
may allow the production of photons with higher energies
(\cite{giommi1998}). Indeed, the absence of any signature of
$\gamma$-ray absorption due to pair production at energies up to 8 GeV
in observations of PKS 2155--304 is consistent with the suggestion that
it contains a relativistic jet oriented nearly along our line of sight
(\cite{vestrand1996}).

\section{Observations}

\subsection{The Mark 6 Telescope}

The University of Durham Mark 6 atmospheric \v{C}erenkov telescope has
been in operation at Narrabri, NSW, Australia since July 1995. The
telescope is described in detail elsewhere (\cite{armstrong1997}). It
uses the well-established imaging technique to separate VHE gamma rays
from the cosmic ray background, combined with a robust noise-free
trigger based on the signals from three parabolic flux detectors of 7 m
diameter and aperture f/1.0 mounted on a single alt-azimuth platform. A
109-element imaging camera ($0.25^\circ$ pixel size) is mounted at the
focus of the central flux detector, with low resolution cameras each
consisting of 19 pixels ($0.5^\circ$ pixel size) mounted at the focus of
the outer (left and right) flux collectors. The telescope is triggered
by demanding a simultaneous temporal (10 ns gate) and spatial
($0.5^\circ$ aperture) coincidence of the \v{C}erenkov light detected in
the three cameras. This triggering system enables the telescope to
detect low energy gamma rays with high immunity from triggering by local
muons. Initial work suggests that our system has a detection probability
for gamma rays of about 1\% at 100 GeV with the probability of detection
rising slowly to about 40\% at 250 GeV.

The Mark 6 telescope has been designed to provide stable operation which
allows the observation of weak DC sources. The multiple-dish triggering
system is stable against variations in performance due to accidental
coincidences. All detector packages are thermally stabilised. The
atmospheric clarity is continuously monitored both using a far infra-red
radiometer and an axial optical CCD camera which enables the position
and visual magnitude of guide stars to be monitored. The gain and noise
performance of the PMTs, digitizer pedestals and associated electronics
are continuously monitored by:

\begin{enumerate}

\item triggering the telescope at random times using a nitrogen laser /
plastic scintillator / optical fibre light guide / opal diffuser system
to simulate \v{C}erenkov flashes and so enable flat-fielding, and

\item producing false triggers at random times to measure samples of the
background noise.

\end{enumerate}

PKS 2155--304 was observed in 1996 September/October/November and 1997
October/November under moonless, clear skies. The 1997 November
observations were made as part of a multiwavelength campaign
co-ordinated by T. Vestrand. An observing log is shown in Table
\ref{observing_log}. Data were taken in 15-minute segments. Off-source
observations were taken by alternately observing regions of sky which
differ by $\pm~15$ minutes in right ascension from the position of PKS
2155--304 to ensure that the on and off segments possess identical
azimuth and zenith profiles. This off-source -- on-source -- on-source
-- off-source observing pattern is routinely used to eliminate any first
order changes in count rate due to any residual secular changes in
atmospheric clarity, temperature etc. The gross counting rate of the
telescope was typically $\sim~800$ per minute near culmination (zenith
angle $5^{\circ}$) reducing to $\sim~150$ per minute at zenith angles of
$60^{\circ}$. Most of the observations at large zenith angles were made
in 1997 November, in order to maximize exposure during the
multiwavelength campaign. Detections of Mrk 501 during its 1997 flaring
activity using the Mark 6 telescope at zenith angles of $> 70^\circ$
have demonstrated the usefulness of TeV observations using the
atmospheric \v{C}erenkov technique at very large zenith angles
(\cite{chadwick1998c}).

\begin{table*}[tb]
\begin{center}
\begin{tabular}{@{}lclc}
\tableline
Date & No. of &Date & No. of \\
& scans & & scans \\
& ON source & & ON source\\ 
\tableline
\tableline

1996 September 3 & 2 & 1997 October 20 & 3 \\
1996 September 5 & 7 & 1997 October 22 & 7 \\
1996 September 6 & 5 & 1997 October 23 & 7 \\
1996 September 7 & 6 & 1997 October 25 & 3 \\
1996 September 8 & 4 & 1997 October 29 & 4 \\
1996 September 10 & 8 & 1997 October 30 & 2 \\
1996 September 12 & 3 & 1997 November 18 & 2 \\
1996 September 14 & 7 & 1997 November 19 & 4 \\
1996 September 15 & 7 & 1997 November 20 & 5 \\
1996 September 17 & 5 & 1997 November 22 & 2 \\
1996 September 18 & 3 & 1997 November 23 & 4 \\
1996 October 9 & 4 & 1997 November 24 & 4 \\
1996 October 10 & 8 & 1997 November 25 & 2 \\
1996 October 12 & 5\\
1996 October 13 & 4\\
1996 October 14 & 6\\
1996 November 2 & 3\\
1996 November 7 & 6\\
1996 November 10 & 4\\
1996 November 11 & 5\\
1996 November 12 & 5\\
\tableline
\end{tabular}
\end{center}

\caption{Observing log for our observations of PKS 2155--304 during
1996 and 1997.}

\label{observing_log} 
\end{table*}

Data were accepted for analysis only if:

\begin{enumerate}

\item the sky was clear and stable, and

\item the gross counting rates in each on-off pair were consistent at
the $2.5~\sigma$ level.

\end{enumerate}

When these criteria are met, we have a total of 41 hours of on-source
observations with an equal quantity of off-source data.

\section{Results}

Routine reduction and analysis of accepted data comprises the following steps:

\begin{enumerate}

\item calibration of the gains and pedestals of all 147 PMTs and digitizer
electronics within a 15 minute segment, using the embedded laser and
false coincidence events,

\item software padding of the data to equalize the effects of on- and
off-source photomultiplier noise on data selection (\cite{cawley1993};
\cite{fegan1997}),

\item identification of the location of the source in the
camera's field of view for each event, using the axial CCD camera,

\item a calculation of the spatial moments of each shower image relative
to the source position, and

\item rejection of events containing an image which would be unlikely to
be produced by gamma rays.

\end{enumerate}

Events considered suitable for analysis are those which are confined
within the sensitive area of the camera (i.e. within $1.1^{\circ}$ of
the centre of the camera) and which contain sufficient information for
reliable image analysis, i.e. which have {\it SIZE} $ > 500$ digital
counts, where 3 digital counts $\sim$ 1 photoelectron, and 200 digital
counts are produced by a 125 GeV gamma ray. 

Monte Carlo simulations indicate that the image shape of the
\v{C}erenkov light from a $\gamma$-ray shower can be approximated by an
ellipse, the major axis of which is oriented towards the source
position, whereas a cosmic ray (hadronic) shower produces a broader,
more irregularly shaped image (\cite{hillas1985}). The image can be
parameterized using techniques developed by the Whipple group which
describe both the shape and the orientation of the image. In addition a
measure of the fluctuations between the centroids of the samples
recorded by the left and right flux collectors of the Mark 6 telescope
provides a further discriminant, $D_{\rm dist}$ (\cite{chadwick1998a}).
Gamma rays are identified on the basis of image shape and left/right
fluctuation, and then plotting the number of events as a function of the
pointing parameter {\it ALPHA}; $\gamma$-ray events from a point source
will appear as an excess of events at small values of {\it ALPHA}.

The selection criteria applied to these data are summarized in Table
\ref{select_table}. They constitute a standard set of criteria developed
to include allowance for the variation of parameters with event size and
are routinely applied to data from all objects recorded at zenith angles
less than $45^{\circ}$. Small changes in the values for {\it WIDTH},
{\it CONCENTRATION} and $D_{\rm dist}$ are made for data recorded at
zenith angles greater than $45^\circ$. These selection criteria differ
little from those used in our earlier reports of the detection of VHE
gamma-rays from PSR B1706--44 (\cite{chadwick1998b}) and Cen X-3
(\cite{chadwick1998a}); a decrease in the weight given to the signal
recorded from the outer (guard ring) detectors in the camera has
produced an {\it ALPHA}-plot for the background with an increasing
population of large {\it ALPHA} events.

% Table 2- parameter selection

\begin{table*}[tb]

\begin{center}

\begin{tabular}{@{}lccccc}

\tableline
Parameter&Ranges&Ranges&Ranges&Ranges&Ranges\\
\tableline
\tableline
{\it SIZE} (d.c.)&$500-800$&$800-1200$&$1200-1500$&$1500-2000$&$2000-10000$\\
{\it DISTANCE}&$0.35^{\circ}-0.85^{\circ}$&$0.35^{\circ}-0.85^{\circ}$&$0.35^{\circ}-0.85^{\circ}$&$0.35^{\circ}-0.85^{\circ}$&$0.35^{\circ}-0.85^{\circ}$\\
{\it ECCENTRICITY}&$0.35-0.85$&$0.35-0.85$&$0.35-0.85$&$0.35-0.85$&$0.35-0.85$\\
{\it WIDTH}&$ < 0.10^{\circ}$&$ < 0.14^{\circ}$&$ < 0.19^{\circ}$&$ < 0.32^{\circ}$&$ < 0.32^{\circ}$\\
{\it CONCENTRATION}&$ < 0.80$&$ < 0.70$&$< 0.70$&$ < 0.35$&$< 0.25$\\
$D_{\rm dist}$&$ < 0.18^{\circ}$&$ < 0.18^{\circ}$&$ < 0.12^{\circ}$&$ < 0.12^{\circ}$&$ < 0.10^{\circ}$\\
\tableline

\end{tabular}

\end{center}

\caption{The image parameter selections applied to the PKS 2155--304
data recorded at zenith angles less than $45^\circ$. For data recorded at zenith 
angles greater than $45^\circ$ small modifications are made to the 
{\it WIDTH, CONCENTRATION} and $D_{\rm dist}$ criteria.} \label{select_table}

\end{table*}

\begin{figure}[tb]

\epsscale{0.5}
\plottwo{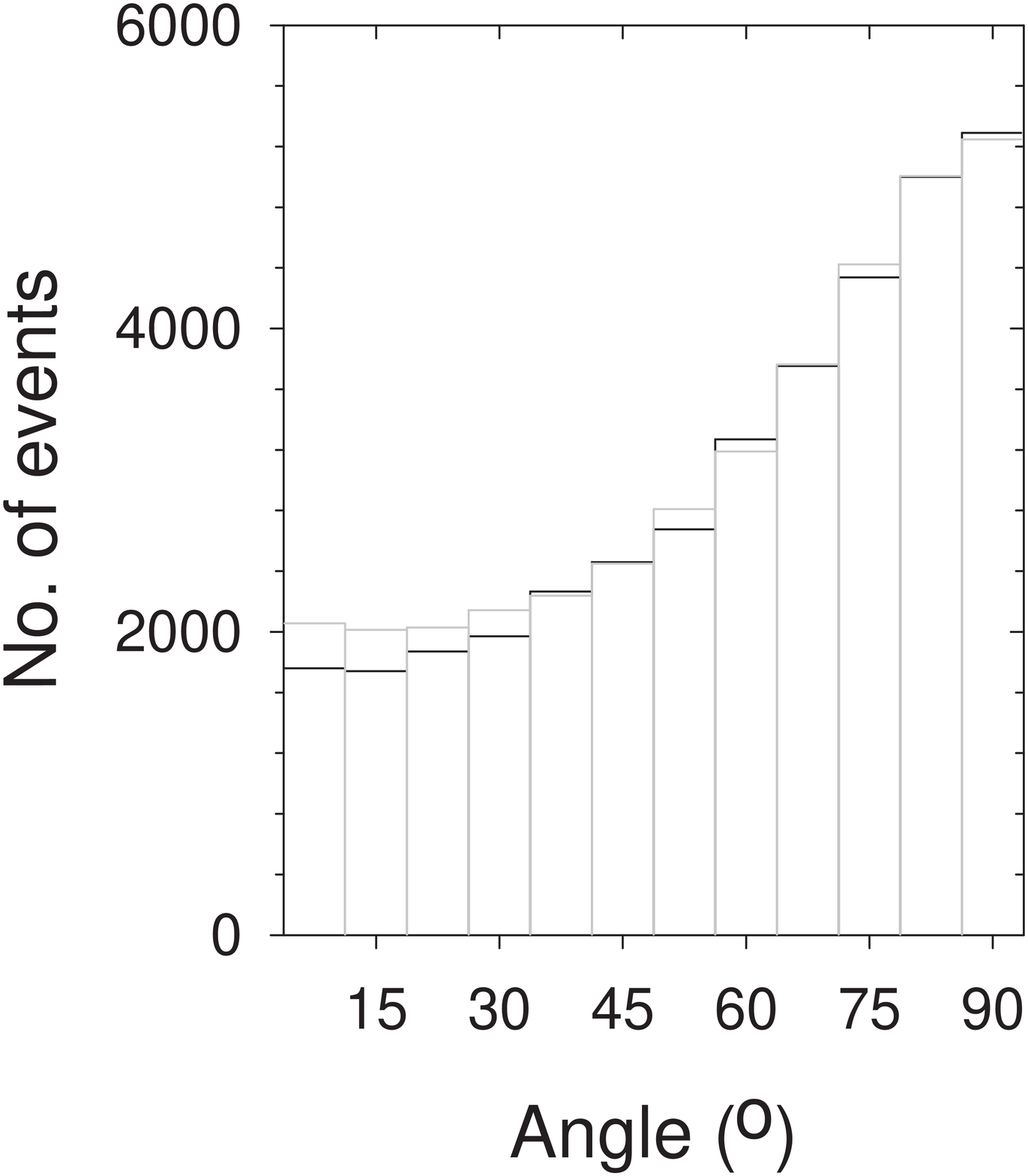}{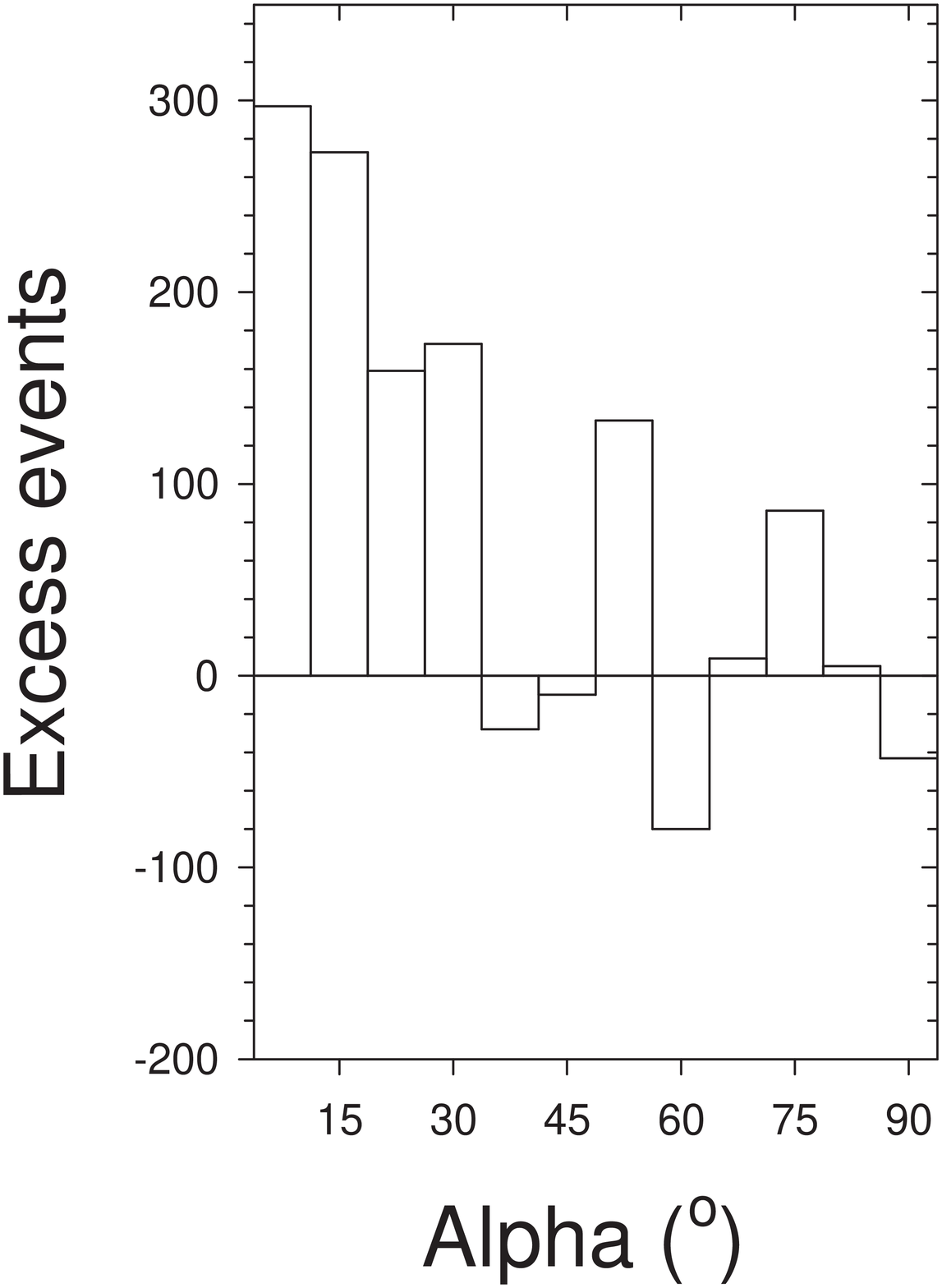}

\caption{(a) The $ALPHA$ distributions ON and OFF source for
PKS 2155--304. The grey line refers to ON source data. (b) The
difference in the {\it ALPHA} distributions for ON and OFF source
events. \label{alpha_plot} }

\end{figure}

The number of events remaining ON and OFF source after the application
of the selections described above are summarized in Table
\ref{result_table}. The {\it ALPHA} distribution for the whole dataset,
after subtracting the results for OFF from ON observations, is shown in
Figure \ref{alpha_plot}, together with the {\it ALPHA} distributions for
the ON and OFF source data individually. No normalization of ON and OFF
data rates has been applied. There is an excess of events at small {\it
ALPHA}, the expected $\gamma$-ray domain, and imposing a selection of
${\it ALPHA} < 22.5^{\circ}$ yields a gamma ray detection significant at
the $6.8~\sigma$ level for the total dataset.

% Table 3 - event selection

\begin{table*}[h]

\begin{center}

\begin{tabular}{@{}lrrrr}

\tableline
& On & Off & Difference & Significance \\
\tableline
\tableline
Number of events & 1021083 & 1023280 & $-2197$ & $-1.5~\sigma$ \\
\\
Number of size and & 600856 & 598733 & 2123 & $1.9~\sigma$ \\
distance selected events & & & & \\
\\
Number of shape & 37125 & 36151 & 974 & $3.6~\sigma$ \\
selected events & & & & \\
\\
Number of shape and & 6099 & 5370 & 729 & $6.8~\sigma$ \\
{\it ALPHA} selected events & & & & \\
\tableline

\end{tabular}

\end{center}

\caption{The results of various event selections for the PKS 2155--304
data. Data from observations at all zenith angles have been combined.}
 
\label{result_table}

\end{table*}

% Fig 1 - Alpha plot

\begin{figure}[tb]
\epsscale{0.5}
\plotone{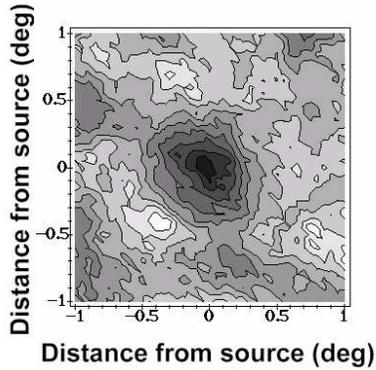}

\caption{The significance of the number of excess events as a function
of assumed source position for PKS 2155--304. The grey scale is such that
black corresponds to a detection probability of $> 6~\sigma$. Contours
are at $0.7~\sigma$ intervals. \label{raster_plot} }

\end{figure}

In addition, a false source analysis has been performed by re-analysing
the ON and OFF-source data using a matrix of assumed source positions in
celestial co-ordinates. For each trial source position in the matrix,
{\it DISTANCE} and {\it ALPHA} are recalculated for each event, and a
selection of events is made with values of {\it DISTANCE} as above and
${\it ALPHA} < 22.5^{\circ}$. The result is shown in Figure
\ref{raster_plot}; the excess of gamma ray candidates maximises when the
source is assumed to be at the centre of the plot. The central point in
the matrix is the position of PKS 2155--304; it is not the geometrical
centre of the camera, which is typically $0.1^{\circ}-0.2^{\circ}$ from
the source position.

\subsection{Observed flux}

544 excess events identified as $\gamma$-rays were detected in 32.5
hours of on-source observation at zenith angles less than $45^{\circ}$.
The current selection procedure for data recorded at zenith angles $ <
45^{\circ}$ is estimated to retain $\sim 20\%$ of the original gamma ray
events. The collecting area has been estimated from Monte Carlo
simulations and is $\sim 5.5 \times 10^{8}~\rm{cm}^{2}$ for these
observations. Using this estimate, the $\gamma$-ray flux incident on the
earth's atmosphere was $(4.2 \pm 0.75_{stat} \pm 2.0_{sys}) \times
10^{-11}~{\rm cm}^{-2}~{\rm s}^{-1}$. The average gamma ray threshold
for this observation, defined as the energy at which the trigger
probability is $\sim 50\%$, was 300 GeV. 

We also note that the high energy (large size) bin for our observations
at zenith angles greater than $45^\circ$ yields a detection significant
at the $ 2 \sigma$ level. We estimate that this sample corresponds to a
minimum $\gamma$-ray energy of about 3 TeV and so there is evidence for
detection of photons with energies $> 3$ TeV from PKS 2155--304.

\subsection{Time variability}

A characteristic of previous observations of VHE gamma rays from XBLs
has been the strong time variability. We have investigated our data for
variations in the TeV flux from PKS 2155--304 on timescales of months
and days.

\begin{figure}[tb]
\epsscale{0.5}
\plotone{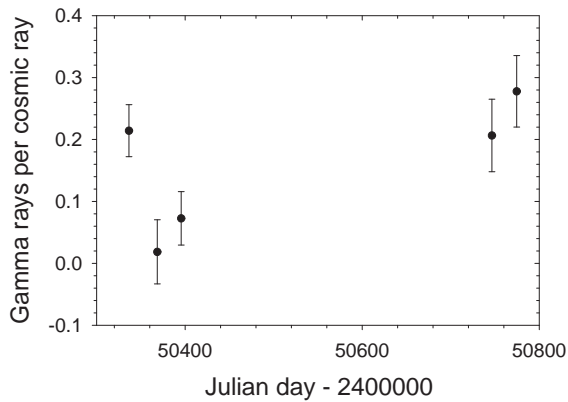}

\caption{The variation of the VHE gamma ray flux from PKS 2155--304
averaged over the observing periods in 1996 September, October and
November and in 1997 October and November.  \label{monthly} }

\end{figure}

We show in Figure \ref{monthly} the variation in the detected
$\gamma$-ray flux averaged over the observations made in the observing
periods in 1996 September, October and November and in 1997 October and
November. The strength of the $\gamma$-ray emission is defined as the
number of $\gamma$-ray candidates (shape and orientation selected
events) to the number of background protons during the same observation.
This makes some allowance for variations in the sensitivity of our
telescope as the zenith angle changes. Using the test for constancy of
emission used by the EGRET group (\cite{mclaughlan1996}), the data
suggest that the VHE $\gamma$-ray emission is time variable. The largest
excursion from the average flux occurred during observations in 1997
November.

\begin{figure}[h]

\epsscale{0.5}

\plottwo{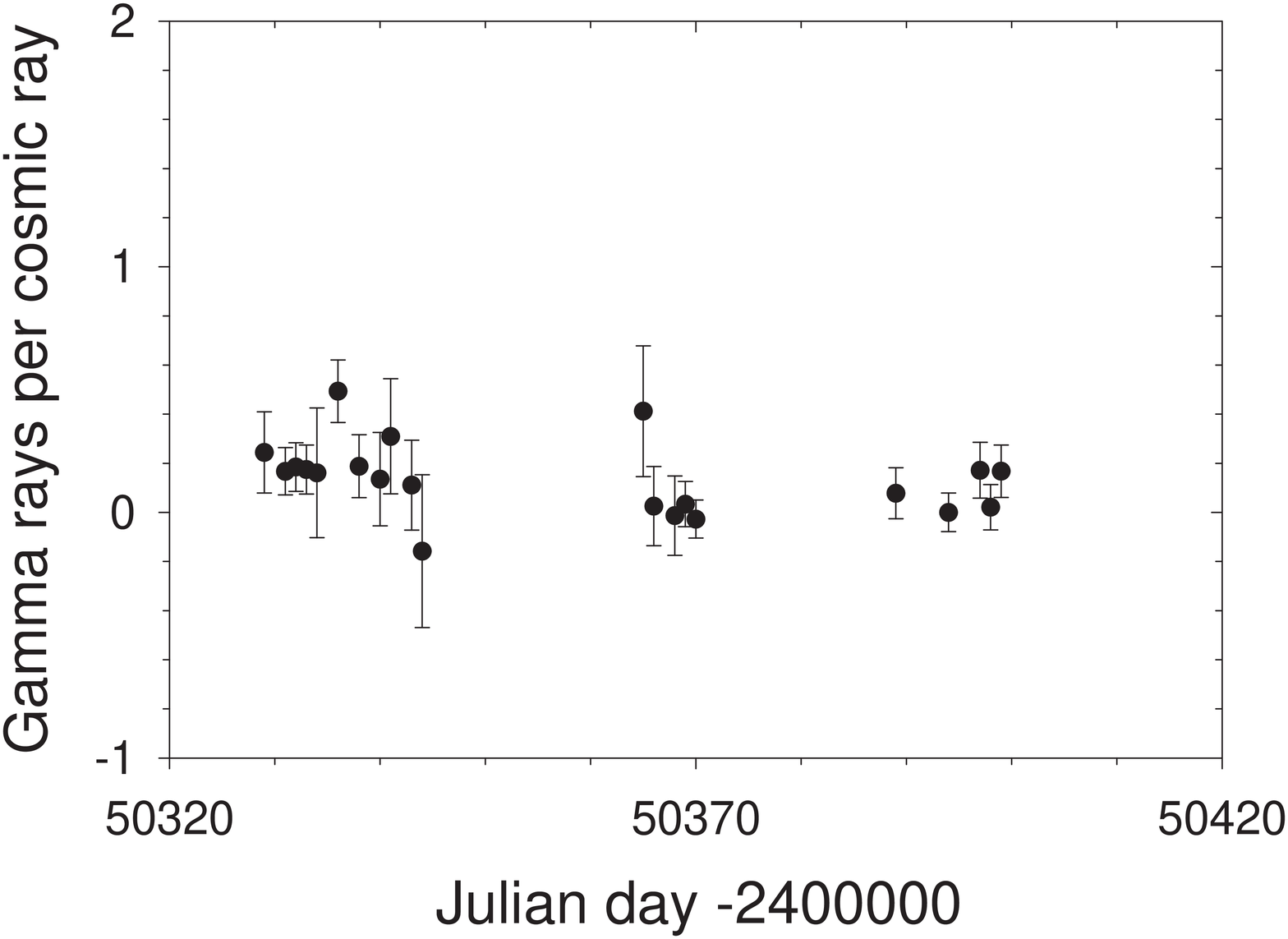}{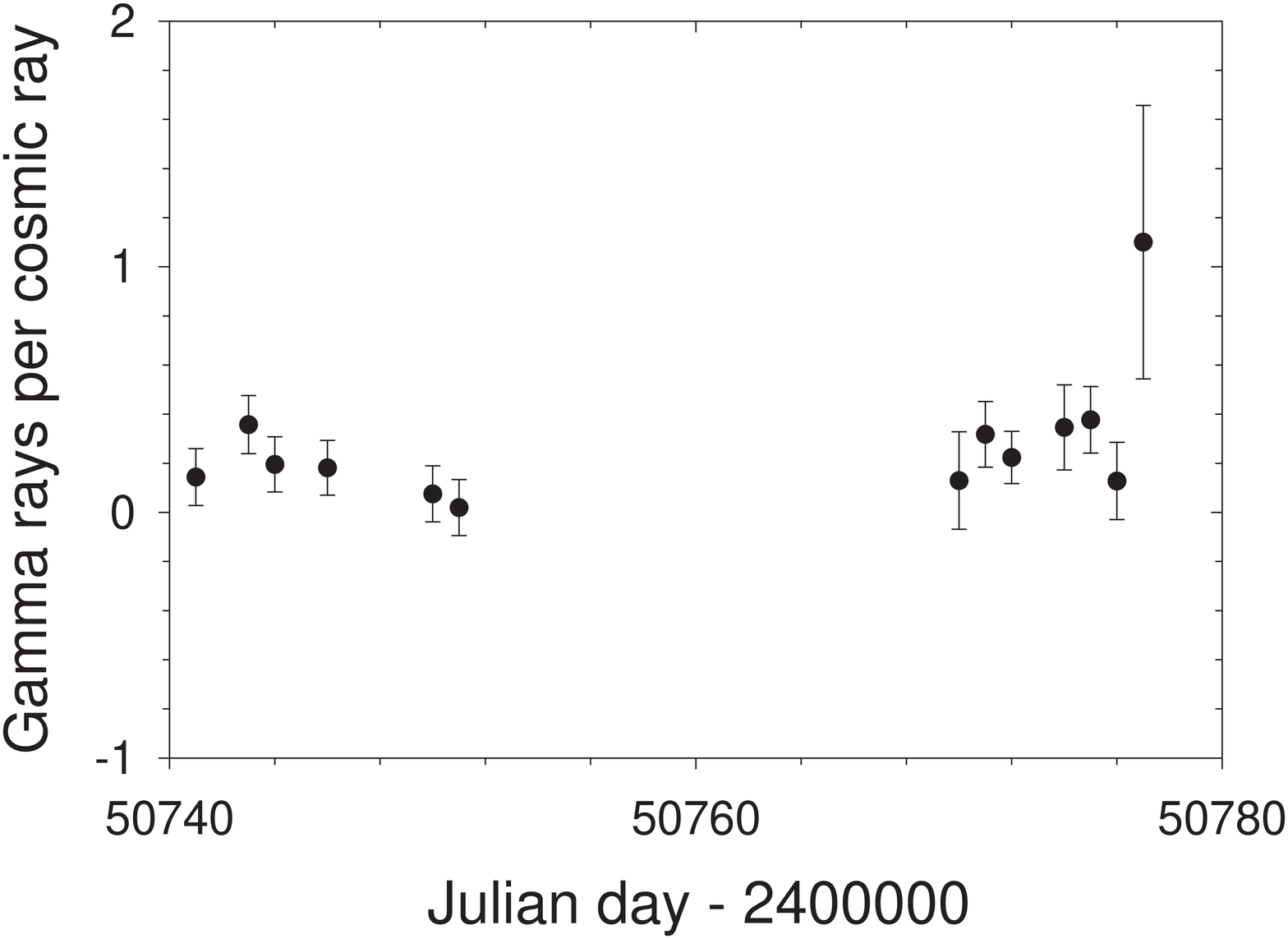}

\caption{The variation of VHE gamma ray flux from PKS 2155--304 from each
observation made in (a) 1996 September, October and November and (b) 1997
October and November. \label{day_to_day} }

\end{figure}

The variation of the gamma-ray emission during observations made in
individual nights is shown in Figure \ref{day_to_day}. We have no
evidence for strong gamma-ray flaring on a time scale of days during
these observations. However, the data suggest that the VHE gamma-ray
emission is variable on a timescale of days, using the test of
\cite{mclaughlan1996}.

\section{Discussion}

PKS 2155--304 is the fourth X-ray emitting BL Lac to be established as a
VHE emitter and, thus far, the furthest from earth with a redshift of
0.117. The emission shows the features of time variability
demonstrated in the other BL Lacs detected at VHE energies. 

\begin{figure}[h]

\centerline{\psfig{file=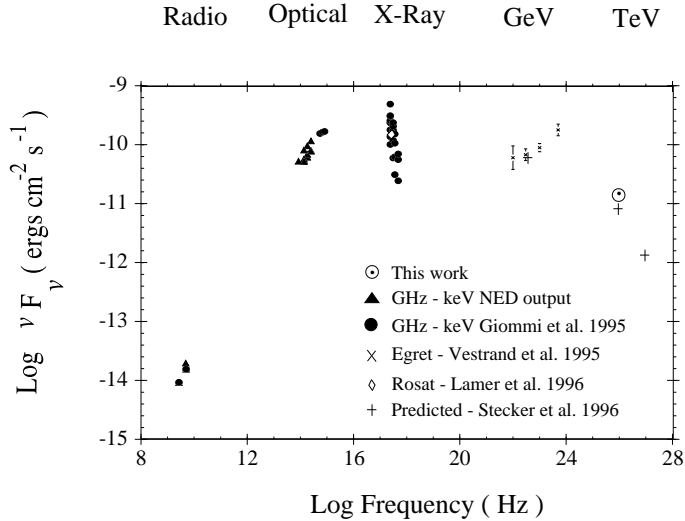,height=10cm,angle=90}}

\caption{The spectral energy distribution of PKS 2155--304. The VHE
point is from the present work. \label{spectrum} }

\end{figure}

In Figure \ref{spectrum} we show the spectral energy distribution (SED)
from PKS 2155--304 from radio to VHE gamma ray energies, including the
present results. The present VHE result is consistent with the SEDs for
the other VHE-emitting blazars. The VHE behaviour is in general
agreement with the predictions of \cite{ghisellini1998} and, as for the
other three established VHE blazars, indicates the success of the
unified model of AGNs in explaining the gross features of VHE emission.
However, the present VHE data are unable to distinguish between the
predictions for the external Compton or the synchrotron self-Compton
model, as is the case for TeV observations of the other VHE blazars. 

% Fig 5 - Spectral energy distribution

\begin{figure}[tb]

\epsscale{0.5}

\plottwo{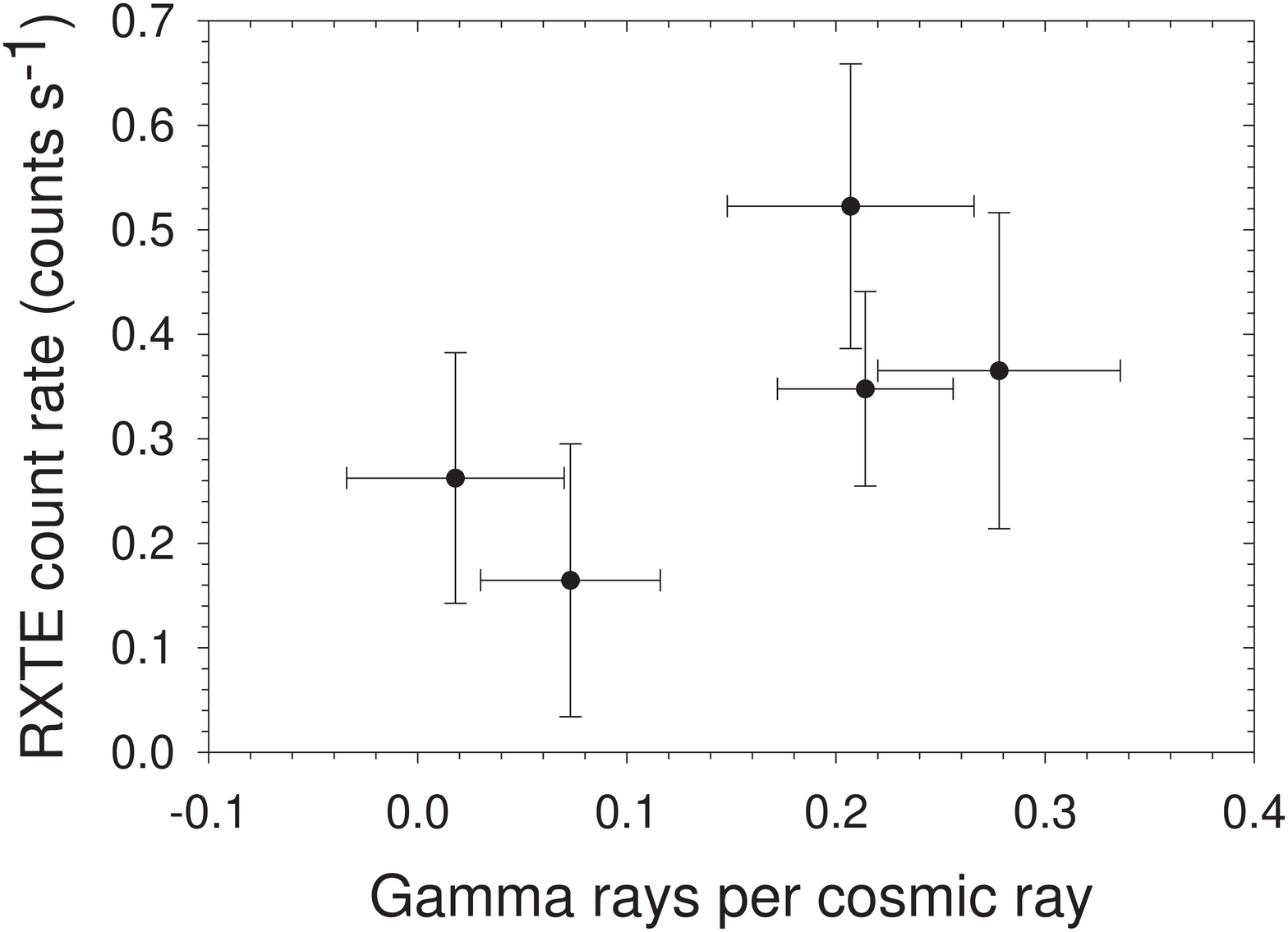}{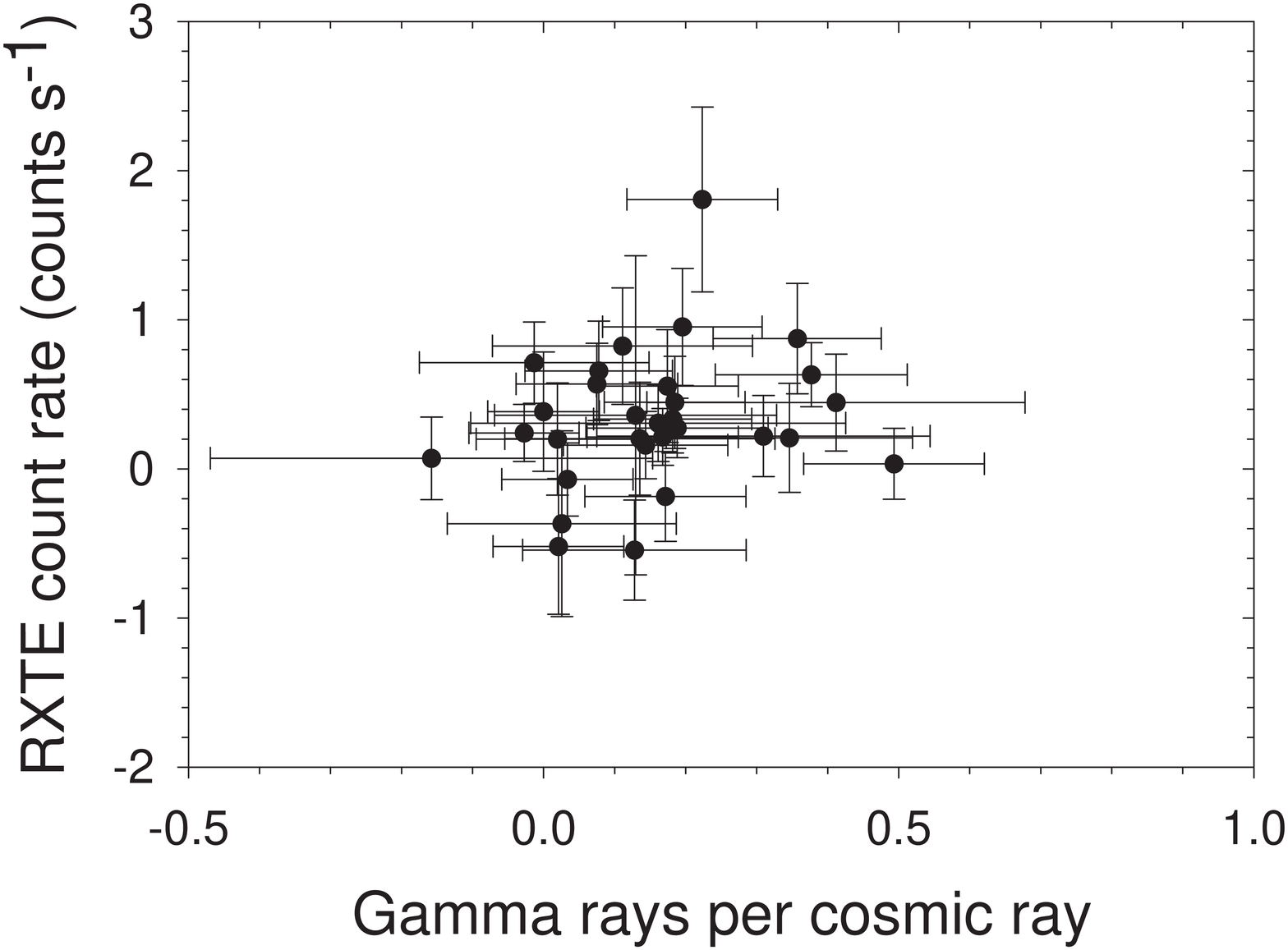}

\caption{The correlation between VHE emission from PKS 2155--304 and the
X-ray flux measured by the {\it RXTE} satellite. Fig 6(a) shows the
correlation when averaged over an observing period ($\sim 10$ days); the
correlation coefficient is $0.66$. Fig 6(b) shows the day-to-day
correlations; the correlation coefficient is $0.22$. \label{xrays} }

\end{figure}

During our observations of PKS 2155--304 measurements of its 2 -- 10 keV
X-ray emission were available from the ASM on {\it RXTE}.
\footnote{Available on the web at
\mbox{http://space.mit.edu/XTE/asmlc/pks2155-304.html}.} In Figure
\ref{xrays}(a) we show the correlation between the average VHE emission
during each of the five months when data were available and the
corresponding average X-ray rate from the {\it RXTE} quick look
analysis. The correlation coefficient for the data is 0.66. A similar
plot for the daily averaged TeV and X-ray measurements are shown in
Figure \ref{xrays}(b). In this case, the correlation coefficient is
0.22.

The data in Figure \ref{xrays} show evidence for a correlation between
the strengths of VHE gamma ray emission and X-ray emission from PKS
2155--304 on timescales of both days and months. In particular, the
observing periods during which we failed to detect significant VHE
gamma-ray emission (1996 October and November) coincide with a period of
low X-ray emission according to the {\it RXTE} ASM measurements. We also
note that \cite{giommi1998} based their conclusions that PKS 2155--304
should not emit VHE gamma rays from an X-ray observation taken with {\it
Beppo-SAX} shortly after our 1996 November (null) observation. The
general correlation between X-ray and VHE emission in PKS 2155--304 is
in agreement with multiwavelength observations of Mrk 421 and Mrk 501
(reviewed by e.g. \cite{ulrich1997}). These show such correlated
behaviour, indicating that the same population of electrons produces the
X-ray and TeV emission; this is the basis for many models for VHE
production in AGNs.

As has been pointed out by \cite{catanese1997a} and \cite{dermer1988}, a
detailed study of the temporal correlation between the X-rays and TeV
gamma-rays during flares can provide a test for beaming. However, none
of the flaring activity observed in our observations of PKS 2155--304 is
at a strength where meaningful limits can be placed on the contribution
of beaming.

The strongest VHE emission from PKS 2155--304 occurred in 1997 November
(during the multiwavelength campaign). We note that the strongest X-ray
emission ever observed from this object was detected by {\it Beppo-SAX}
during observations in 1997 November 22 -- 24 (\cite{chiappetti1997}).
EGRET also detected GeV gamma-rays, again at a flux considerably higher
than prevous detections, during an observation from 1997 November 11 --
17, just before the onset of our November observation
(\cite{sreekumar1997}). The results of the 1997 November multiwavelength
campaign will be reported elsewhere.

Our observations indicate that PKS 2155--304 emits VHE gamma rays at
energies above about 300 GeV and that the energy spectrum extends, with
significance at the $2 \sigma$ level, beyond 3 TeV. The estimated
average integral flux above a threshold of 300 GeV is $( 4.2 \pm
0.7_{stat} \pm 2.0_{sys}) \times 10^{-11}~{\rm cm}^{-2}~{\rm s}^{-1}$.

Observations of VHE gamma-rays from AGNs can be used to probe the
inter-galactic infra-red radiation field (\cite{stecker1992}). The most
recent version of this model (\cite{stecker1997b}, \cite{stecker1998})
uses estimates of the infra-red radiation field due to \cite{malkan1998}
to predict the optical depth for TeV sources as a function of distance.
They predict an optical depth of 1 -- 2 for an object emitting 3 TeV
gamma-rays at $ z = 0.12$. Our detection of $> 3$ TeV gamma-rays from PKS
2155--304 is thus not in conflict with the predictions of
\cite{stecker1998}. However, a more sensitive test must await a
determination of the VHE gamma-ray spectrum from our data.

\acknowledgments

We are grateful to the UK Particle Physics and Astronomy Research
Council for support of the project and the University of Sydney for the
lease of the Narrabri site. The Mark 6 telescope was designed and
constructed with the assistance of the staff of the Physics Department,
University of Durham. GJW acknowledges the receipt of a Tempus
studentship. The efforts of Mrs. S.E. Hilton and Mr. K. Tindale are
acknowledged with gratitude. This paper uses quick look results provided
by the ASM/{\it RXTE} team and uses the NASA/IPAC Extragalactic Database
(NED) which is operated by the Jet Propulsion Laboratory, Caltech, under
contract with the National Aeronautics and Space Administration.

\end{document}